\documentclass[aps,twocolumn,amsmath,amssymb,showpacs,floatfix]{revtex4}
\usepackage{graphicx}

\begin{document}
\title[Zone-plate focusing and lithography using Bose-Einstein Condensates]{Zone-plate focusing of Bose-Einstein condensates for atom optics and erasable high-speed lithography of quantum electronic components}
\author{T.E. Judd$^{1,2}$, R.G. Scott$^1$, G. Sinuco$^1$, T.W.A. Montgomery$^1$, A.M. Martin$^3$, P. Kr{\"u}ger$^1$, and T.M. Fromhold$^1$}
\address{$^1$Midlands Ultracold Atom Research Centre, School of Physics \& Astronomy, University of Nottingham, Nottingham NG7 2RD, United Kingdom \\
$^2$Physikalisches Institut, Eberhard-Karls-Universit{\"a}t T{\"u}bingen, CQ Center for Collective Quantum Phenomena and their Applications, Auf der Morgenstelle 14, D-72076 T¨ubingen, Germany\\
$^3$School of Physics, University of Melbourne, Parkville, Vic. 3010, Australia}
\date{\today}

\begin{abstract}
We show that Fresnel zone plates, fabricated in a solid surface, can sharply focus atomic Bose-Einstein condensates that quantum reflect from the surface or pass through the etched holes. The focusing process compresses the condensate by orders of magnitude despite inter-atomic repulsion. Crucially, the focusing dynamics are insensitive to quantum fluctuations of the atom cloud and largely preserve the condensates' coherence, suggesting applications in passive atom-optical elements, for example zone plate lenses that focus atomic matter waves and light at the same point to strengthen their interaction. We explore transmission zone-plate focusing of alkali atoms as a route to erasable and scalable lithography of quantum electronic components in two-dimensional electron gases embedded in semiconductor nanostructures. To do this, we calculate the density profile of a two-dimensional electron gas immediately below a patch of alkali atoms deposited on the surface of the nanostructure by zone-plate focusing. Our results reveal that surface-induced polarization of only a few thousand adsorbed atoms can locally deplete the electron gas. We show that, as a result, the focused deposition of alkali atoms by existing zone plates can create quantum electronic components on the 50 nm scale, comparable to that attainable by ion beam implantation but with minimal damage to either the nanostructure or electron gas. 
\end{abstract}


\maketitle
\section{Introduction}
Cooling alkali atoms to $\mu$K temperatures and below has opened the field of atom optics, leading to many breakthroughs in both fundamental physics and emerging applications \cite{cronin}. It provides new ways to control the atoms, by tailoring their potential landscape on a $\mu$m scale \cite{cronin,andreas1,judd,pasquini,pasquini2,oberst}, probe their environment, for example in high-precision matter-wave sensors \cite{Wild1,DeKieviet} or atomic microscopes \cite{carnal,doak,doak2}, and use them for matter-wave lithography of nanostructures \cite{McClelland,Timp1,Timp2,ODwyer,pfau}. Focusing the atomic matter waves is crucial to the development of such instruments, and for emerging applications involving the transfer of cold atoms into hollow-core optical fibres \cite{christensen,hofferberth}, but remains a challenging task. Usually, spot focusing is achieved by using electromagnetic lenses \cite{fallani}, which requires sophisticated equipment, or by reflecting the atoms from curved optical \cite{bongs,Berkhout} or magnetic mirrors \cite{arnold, Merimeche,Saba}, which are hard to make and keep atomically clean \cite {doak,Holst}. To avoid these complications, matter waves can be focused by diffracting them from commercially-available Fresnel zone plates (ZPs) \cite{carnal,doak,doak2}, comprising a series of concentric circular apertures, whose focal length is proportional to the speed of the incident atoms \cite{meyer}. So far, though, ZP focusing of atomic matter waves has only been demonstrated for non-interacting He or Ne atoms with approach speeds $\gtrsim$ 400 m s$^{-1}$, corresponding to long focal lengths $\gtrsim$ 30 cm \cite{carnal,doak,doak2}.

Here, we show that such ZPs can sharply focus Bose-Einstein condensates (BECs) that are partially transmitted through the plate or, if they approach it slowly enough ($\sim 1-10$ $\text{mm s}^{-1}$), quantum reflect from the attractive Casimir-Polder (CP) atom-surface potential \cite{pasquini,pasquini2,scott,scottnoise,cornish}. We use numerical solutions of the Gross-Pitaevskii equation, supplemented by a truncated Wigner approach \cite{norrieturb,scottnoise}, to investigate how the focusing dynamics depend on the ZP geometry and on the parameters and incident velocity of the BEC. Our calculations reveal that the focal length is similar to that expected from a single-particle ray picture but, for sufficiently high atom densities, also exhibits resonances that originate from inter-atomic interactions. At the low approach speeds required for quantum reflection to occur, the BECs focus within 100\:$\mu\text{m}$ of the surface, where trapping usually occurs in atom chips \cite{hinds,Folmanrev,Reichelrev,fortaghrev}. Since, in this regime, the deBroglie wavelength of the incident BEC is comparable with that of light, a single ZP can focus both atoms and light \emph{at the same point}, thereby promoting the strong light-matter interaction required for few-photon nonlinear optics \cite{kimble,hofferberth,lightfocus}. Focusing is sharpest for dilute pancake-shaped BECs, which have a narrow distribution of approach speeds and hence exhibit little chromatic aberration \cite{meyer}, are less susceptible to disruption by dynamical excitations created during interaction with a surface \cite{scott,scottnoise}, span many rings of the ZP so that its resolution is intrinsically high \cite{meyer}, and experience less defocusing by inter-atomic repulsion. Despite this repulsion, ZP focusing can increase the BEC's density by orders of magnitude, raising the possibility that quantum fluctuations will significantly reduce the condensate fraction \cite{scottnoise}. Surprisingly, though, we find that such fluctuations cause little depletion of the BEC. 

Finally, we introduce ZP focusing of matter waves onto a semiconductor nanostructure as a route to erasable and scalable nm-precision lithography of quantum electronic components fabricated in a two-dimensional electron gas (2DEG) on, or just below, the surface. Erasable lithography is of great interest for studying quantum transport and control, but has so far been achieved only in a small number of laboratories using scanning probe techniques \cite{crook,crookref12}. Alkali atoms deposited on a semiconductor surface polarize because their valence electron partially transfers to the surface \cite{polarize,polarize2}. We show that this repels 2DEG electrons strongly enough to produce local insulating regions with dimensions determined by the focal width of the BEC, which can be made $\lesssim$ 50 nm using existing ZPs. Compared with existing fabrication techniques, ZP lithography using matter waves offers several key advantages, which we identify and discuss. 

The paper is organised as follows. In Section \ref{sec:Quantum reflection from the zone plate}, we define the system parameters, introduce our model for calculating the BEC dynamics, and use this model to study quantum reflection focusing from a ZP etched in a Si surface. In Section \ref{sec:Transmission focusing and decoherence of the BEC}, we investigate focusing induced by transmission through, and quantum reflection from, a free-standing ZP structure and determine how the focusing process affects the quantum coherence of the atom cloud. In Section \ref{sec:Zone-plate lithography of two-dimensional electron gases}, we explore transmission ZP focusing as a route to fabricating erasable quantum components in high-mobility electron gases. We summarize our results and draw conclusions in Section \ref{sec:Summary}.   

\section{Quantum reflection from a zone plate}
\label{sec:Quantum reflection from the zone plate}

In this section, we consider quantum reflection of a $^{23}$Na BEC from a ZP etched in a Si surface. This system produces a single focus, which is sharp due to the narrow velocity distribution of atoms within the BEC, and is therefore simple enough to elucidate the key features of the focusing dynamics, in particular the effect of inter-atomic interactions.

When an alkali atom approaches within $\sim 3$ $\mu$m of a planar surface, its potential energy decreases rapidly due to mutual polarization of the atom and the surface \cite{shimizu,pasquini,pasquini2,scott}. At distance $x'$ from a perfectly conducting surface, the atom-surface interaction can be described by the CP potential energy $V_{CP}(x')=-C_4/{x'} ^{3}(x' + 3\lambda_a / 2\pi^2)$, where, for a Na atom, $C_4=9.1\times10^{-56}$ Jm$^{4}$, and $\lambda_a=590$ nm is the effective atomic transition wavelength \cite{pasquini}. This attractive potential creates no classical turning point for an incident atom. But if the atom's approach velocity, $v_x$, is sufficiently low, $\sim 1-10$ $\text{mm s}^{-1}$, the corresponding deBroglie wavelength, $\lambda_{dB}$, is long enough to span the rapidly decreasing CP potential, causing quantum reflection to occur for both non-interacting atoms \cite{shimizu,shimizu2,friedrich,heller} and BECs \cite{pasquini,pasquini2}. For BECs, quantum reflection probabilities up to 0.7 have been achieved by etching an array of $\sim 100$ nm-diameter pillars into the surface to enhance the action of the CP potential \cite{pasquini2}. In these experiments, the spacing of the pillars ($0.5$ $\mu$m) was chosen to be $\ll \lambda_{dB}$ to avoid diffracting the matter waves. By contrast, here we study matter waves scattering from a larger, $100$\:$\mu$m scale, ZP structures specifically designed to diffract, and hence focus, an incident BEC. 

\begin{figure}
\includegraphics[width=1.0\columnwidth]{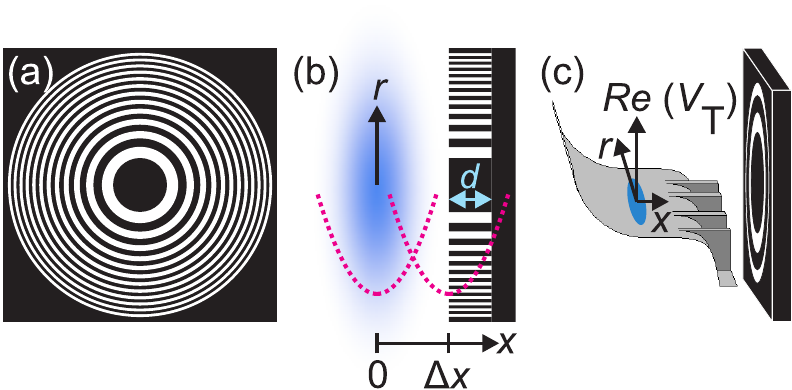} 
\caption{\label{fig:fullsys}(a) Plan view of the ZP showing raised (black) and etched (white) rings. (b) Black shape: schematic cross-section in the $x-r$ plane (axes inset) through the ZP, etched to depth $d$ (horizontal arrow). Blue shape: initial position of the BEC when the harmonic trap is centred at $x=0$. Left (right) dotted parabola: harmonic trap potential before (after) displacement. (c) Gray surface: real part of potential energy $V_{T}(x,r)$, defined in the text, shown for $x<\Delta x + d$ (i.e. above the top surface of the ZP or within an etched ring) after displacement of the harmonic trap. The displaced trap accelerates the BEC (blue) towards the ZP (right). In (c), few ZP rings are shown for clarity.}
\vspace{-0.4cm}
\end{figure}

We first consider a ZP comprising 12 concentric rings etched, by standard plasma etching techniques, for example, to a depth $d=$ 10\:$\mu$m into a flat Si surface, which lies in the $x = \Delta x$ plane [Fig. \ref{fig:fullsys}(a,b)]. Note that $\Delta x$ is a variable that we change in order to control the initial position of the BEC relative to the surface (details below). The ring edges are at $r = \sqrt{y^{2}+z^{2}} = R_n=R_1 n^{1/2}$ ($n=1,2,3,4,...,24$), where $R_1$ = 20\:$\mu$m is the radius of the inner raised disk [black in Fig. \ref{fig:fullsys}(a)]. In Fig. \ref{fig:fullsys}(b), the black shape shows a schematic cross-section through the ZP, with the etched rings appearing as white indentations. 

For a single Na atom of mass $m$, represented by a plane wave of wavelength, $\lambda_{dB}$, travelling at velocity, $v_x$, along the $x$-axis, simple ray analysis \cite{meyer} predicts that constructive interference between wavefronts that quantum reflect from adjacent raised rings will focus the wave at a distance 
\begin{equation}
\label{eq:focus}
f=R{_1}{^2}/\lambda_{dB} = R{_1}{^2}mv_x/h = 0.023v_x
\end{equation}
from the ZP, provided $f \gg$ the outer radius, $R_{24}$ = $98$\:$\mu$m, of the largest ZP ring. The wave can still focus if this condition is not satisfied, but the focal length will differ from that predicted by Eq. (\ref{eq:focus}).

To investigate how a BEC interacts with the ZP, we first consider the geometry of the system. Since the BEC approaches the ZP along a common ($x$) axis of circular symmetry, we describe the system using cylindrical, $(x,r)$, co-ordinates [Fig. \ref{fig:fullsys}(b,c)]. For each atom, the ZP creates an attractive CP potential energy, $V_{zp}(x,r)$. If $r<R_1$ or $R_{2n}\leq r \leq R_{2n+1}$, so that the atom is directly above one of the raised rings, we take this potential energy to be 

\label{eq:zpetchedv} 
\begin{displaymath}
V_{zp}(x,r)=\left\{
\begin{array}{l}
V_{CP}(x'),\:\:\:\:\:\:\:\:\:\:\:\:\:\:\:\:\:\:\:\:\:\:\:\:\:\:\:\:\:\:\:\:\:\:\:\:\:\:\:\:\:\:\: \rm{if} \; \emph{x}'>\delta \\ 
V_{CP}(\delta) -  i(x - \Delta x + \delta)V_{im}\:\:\:\:\:\:\: \rm{if} \; \emph{x}'\leqslant \delta,
\end{array}
\right.  
\end{displaymath}
where $x'=\Delta x -x$ is the atom-surface separation and $V_{im}=1.6\times10^{-26}$ Jm$^{-1}$ \cite{scott}. The complex form of $V_{zp}$, within distance $\delta=0.15$ $\mu$m of the surface, avoids the divergence of $V_{CP}(x')$ as $x'\rightarrow 0$ and models adsorption of those atoms that reach the surface \cite{scott}. If $R_{2n-1}< r < R_{2n}$, so that the atom approaches an {\em{etched}} ring, we take $V_{zp}(x,r)=0$ if $x<\Delta x + d$ [i.e. above the surface or within the etched ring: see Fig. \ref{fig:fullsys}(b)] and, to simulate the adsorption of atoms that enter the etched ring \cite{footnote8}, $V_{zp}(x,r)=-  i[x - (\Delta x + d)]V_{im}$ if $x\geqslant \Delta x + d$ (i.e. beyond the bottom of the ring).

To study quantum reflection from the ZP, we adapt the system used in recent experimental observations of quantum reflection for a BEC approaching a planar Si surface \cite{pasquini,pasquini2}. We first consider a dilute condensate, henceforth called BEC $A$, containing $N=3\times 10^5$ $^{23}$Na atoms in a harmonic trap with cylindrical symmetry about the $x$-axis and frequencies $\omega_x=2\pi\times$3.3 rad s$^{-1}$ and $\omega_r=2\pi\times$1.0 rad s$^{-1}$  in the longitudinal ($x$) and radial ($r$) directions respectively. This creates a pancake-shaped BEC [Fig. \ref{fig:fullsys}(b)] with longitudinal width $l_x \approx 40$ $\mu\text{m}$, radial diameter $l_r \approx 180$ $\mu\text{m}$, and peak density $n_0=6.3\times 10^{17}$ m$^{-3}$.  We choose the pancake shape to limit disruption of the BEC during quantum reflection \cite{scott,scottnoise}. 

Initially, the BEC is in its equilibrium ground state, centred at $x=r=0$. At time $t=0$, we suddenly displace the harmonic trap by a distance $\Delta x$ along the $x$ axis, so that its centre coincides with the top surface of the ZP at $x= \Delta x$ [Fig. 1(b)]. This causes the BEC to approach the ZP with velocity $v_{x}\thickapprox \omega_x\Delta x=\overline{v}_x$ at time $T\approx \pi/2\omega_{x}$. We consider $\Delta x$ values for which $\overline{v}_x \geqslant 2$ mm s$^{-1}$ to avoid creating dynamical excitations during the reflection process \cite{pasquini,pasquini2,scott}. After the trap displacement, the total potential energy of each Na atom in the BEC is $V_T(x,r)=V_{zp}(x,r)+\frac{1}{2}m\left[\omega_{x}^{2}(x-\Delta x)^2+\omega_{r}^{2}r^{2}\right]$. As shown in Fig. 1(c), the real part of $V_{T}(x,r)$ decreases rapidly near the top surface of the ZP, but is constant within the etched rings \cite{footnote8}. At time $T$, we switch off the harmonic trap to prevent it influencing the subsequent focusing process. We determine the dynamics of the BEC by using the Crank-Nicolson method \cite{scott} to solve the Gross-Pitaevskii equation
\begin{equation}
\label{eq:gpe}
i\hbar\frac{\partial\Psi}{\partial t}=-\frac{\hbar^2}{2m}\nabla^2\Psi+V_{T}\Psi+\frac{4\pi\hbar^{2}a}{m}\left|\Psi\right|^2\Psi,
\end{equation}
where $a=2.9$ nm is the $s$-wave scattering length for Na, $\nabla^2$ is the Laplacian in cylindrical coordinates, and $\Psi(x,r,t)$ is the wave function at time $t\geqslant0$, normalized so that $\left|\Psi\right|^2$ is the number of atoms per unit volume.

\begin{figure}
\includegraphics[width=1.\columnwidth]{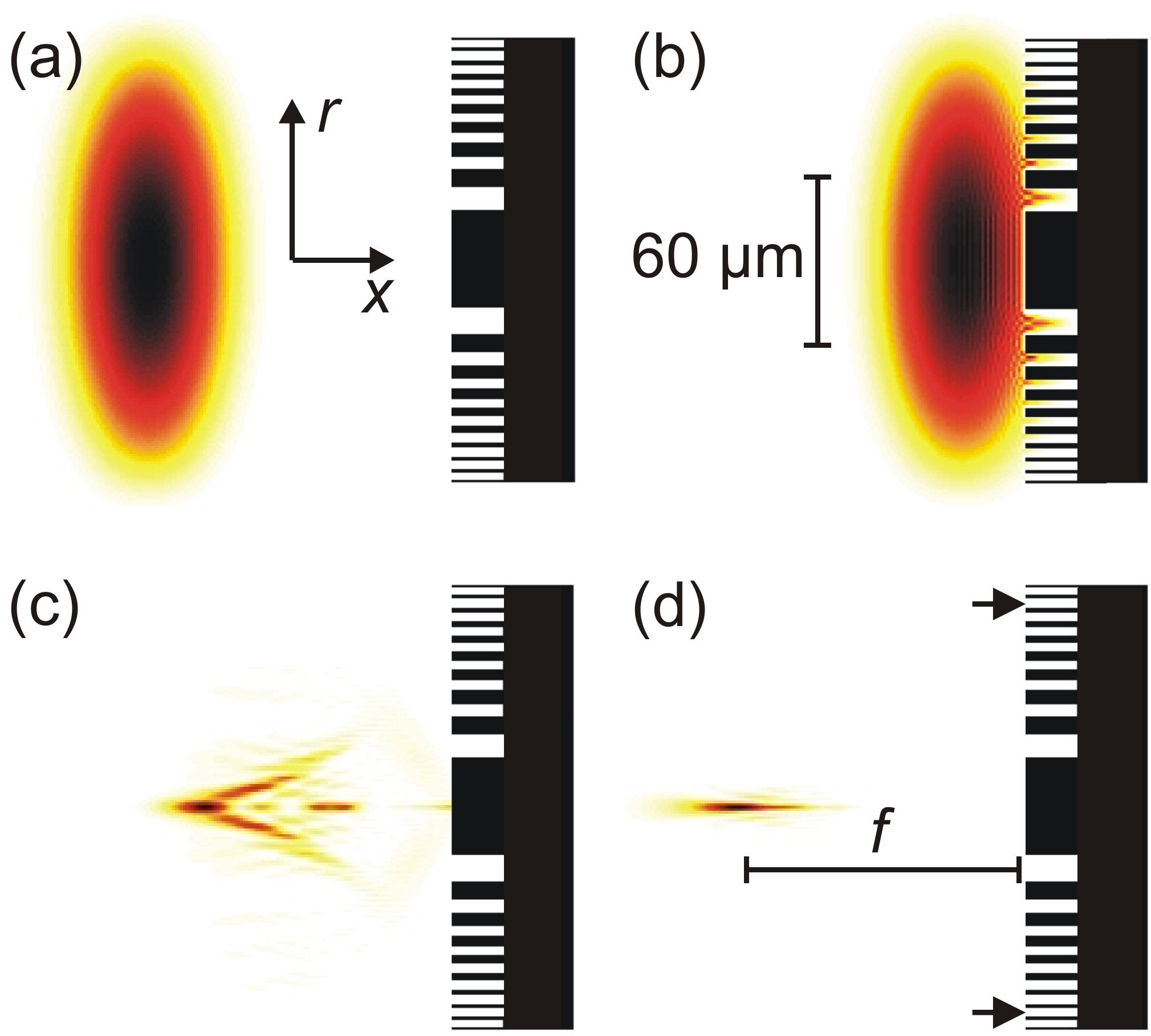} 
\caption{\label{fig:basic}Atom density (dark high) in the $x-r$ plane (axes inset) at $t$ = 0 (a), 72 ms (b), 90 ms (c), and 99 ms (d) for $\Delta x = 240$ $\mu$m $(\overline{v}_x=5$ mm\:s$^{-1})$. Black shapes are schematic cross-sections through the ZP. Vertical bar in (b) indicates scale. In (d), horizontal bar shows the focal length \cite{footnote7} and arrows mark the narrowest ring spanned by BEC $A$.}
\vspace{-0.4cm}
\end{figure}
Figure \ref{fig:basic} shows atom density profiles at key stages during the quantum reflection and focusing of BEC $A$ following a trap displacement $\Delta x = 240$ $\mu$m $(\overline{v}_x=5$ mm s$^{-1})$. Immediately after the trap displacement, the BEC remains centred at $x=0$ [Fig. \ref{fig:basic}(a)], but accelerates towards the ZP. At $t=$ 72 ms [Fig. \ref{fig:basic}(b)], the BEC's leading edge has reached, and undergone partial quantum reflection from, the rapidly-decreasing CP potential near the raised rings. Interference between the incident and reflected matter waves weakly modulates the atom density profile [vertical red and black stripes in Fig. \ref{fig:basic}(b)]. In addition, some atoms have entered the etched rings. By $t=$ 90 ms [Fig. \ref{fig:basic}(c)], the reflected atoms have moved away from the ZP and formed an ``arrow head'' density pattern. The upper and lower edges of the arrow head approach one another, moving towards $r=0$ where they meet, and transiently focus, at $t=$ 99 ms [Fig. \ref{fig:basic}(d)] before diverging again \cite{footnote7}. It might be possible to achieve similar compression of a BEC by using a scanning focused laser beam, rather than a micro-fabricated diffraction grating like that in Fig. \ref{fig:fullsys}, to imprint the ZP pattern optically on the atom cloud \cite{opticalZP}.

Comparison of Figs. \ref{fig:basic}(a) and (d) reveals that the width of the focused BEC along the $x$ axis is similar to that of the initial state because atoms at the front of the BEC reflect and focus before those at the back. Consequently, the size of the focused BEC can be reduced by decreasing $l_x$. As expected from both ray and wave analyses \cite{meyer,ZP_width_theory}, the radial width, $l^{f}_{r}$, of the focused cloud approximately equals the width of the narrowest ring that the atoms enter. Atoms can only enter the $j^{th}$ etched ring if their incident momentum exceeds that of the lowest quantized radial mode in the ring, which requires $\overline{v}_x\geq v_j=h/2mw_j$, where $w_j=(R_{2j}-R_{2j-1})$ is the ring width \cite{atom_resonance,footnote10}. Resonant injection of atoms into the narrowest ($8^{th}$) ring spanned by BEC $A$, marked by arrows in Fig. \ref{fig:basic}(d), only occurs if $\overline{v}_x\geq v_8=3.4$ mm s$^{-1}$. In this regime, $l^{f}_{r}\thickapprox w_8=2.5$ $\mu$m [Fig. \ref{fig:basic}(d)] and so the volume of the focused cloud is a factor $\thickapprox(l^{f}_{r}/l_r)^2=1.9\times 10^{4}$ smaller than the initial BEC. As $\overline{v}_x$ decreases, the width of the narrowest ring that the BEC can penetrate gradually increases, causing $l^{f}_{r}$ to increase approximately as $1/\overline{v}_x$.

\begin{figure}[t]
\includegraphics[width=1.\columnwidth]{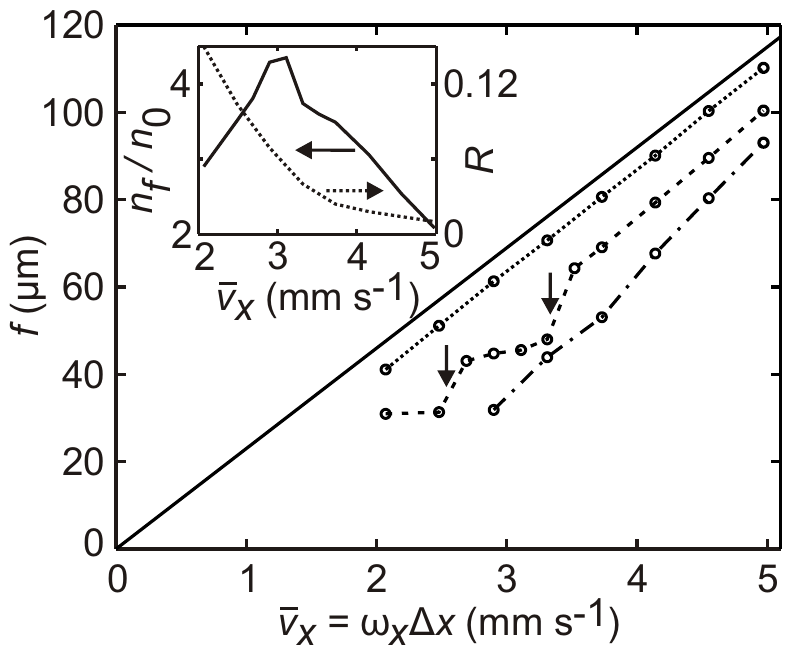} 
\caption{\label{fig:focal}$f(\overline{v}_x)$ curves calculated from Eq. (\ref{eq:focus}) (solid line) and from numerical solutions of Eq. (\ref{eq:gpe}) for a single-atom wavepacket (dotted curve), BEC $A$ (dashed curve) and BEC $B$ (dot-dashed curve) after quantum reflection focusing. Symbols show data points. Vertical arrows mark $\overline{v}_x$ values where $f$ changes abruptly, as explained in the text. Inset: curves showing $n_f/n_0$ (solid) and $R$ (dotted) calculated versus $\overline{v}_x$ for BEC $A$.}
\vspace{-0.4cm}
\end{figure}

We now investigate how $f$, and the underlying focusing dynamics, vary with $\overline{v}_x$. The solid curve in Fig. 3 shows $f(\overline{v}_x)$ calculated from Eq. (\ref{eq:focus}) for a single Na atom modeled by an incident plane wave \cite{footnote7}. If the atom is, instead, described by a wavepacket, using Eq. (\ref{eq:gpe}) with $a=0$, we obtain the dotted $f(\overline{v}_x)$ curve in Fig. \ref{fig:focal} \cite{footnote2}. This curve lies slightly below the solid line given by Eq. (\ref{eq:focus}) primarily because the assumption that $f \gg R_{24} = 98$ $\mu$m made in deriving Eq. (\ref{eq:focus}) is not strictly valid \cite{meyer}.

In Fig. \ref{fig:focal}, the $f(\overline{v}_x)$ curve calculated for BEC $A$ (dashed curve) reveals that inter-atomic interactions further reduce $f$. As the BEC starts to quantum reflect, atoms accumulate near the ZP surface and their repulsive mean field decelerates those atoms that are still approaching the ZP, thus reducing the BEC's mean incident velocity and, consequently, also reducing $f$. Resonant injection into the ZP's rings reduces the build up of atoms and decelerating mean field potential near the entrance to the ring, causing the approach speed and $f$ to increase abruptly with increasing $\overline{v}_x$, as indicated by the two vertical arrows in Fig. \ref{fig:focal}. The exact $\overline{v}_x$ values at which these resonances occur depend on the mean field inter-atomic repulsion within the rings \cite{footnote10}, which varies rapidly in space and time during the reflection process. Consequently, a simple non-interacting model can only estimate the positions of the resonances. However, the two large abrupt changes in $f$ marked by the vertical arrows in Fig. \ref{fig:focal} occur close to the $\overline{v}_x$ required for resonant injection into the single-particle modes of two rings simultaneously. Specifically, the resonant feature indicated by the left-hand (right-hand) arrow appears to originate from co-excitation of the lowest and first excited radial modes of the 3$^{rd}$ and 1$^{st}$ (7$^{th}$ and 2$^{nd}$) rings respectively.

The dot-dashed $f(\overline{v}_x)$ curve in Fig. \ref{fig:focal} is calculated for a denser condensate, BEC $B$, comprising $N = 3\times 10^6$ $^{23}$Na atoms in a harmonic trap with $\omega_x(\omega_r)=2\pi\times$9.9 (3.0) rad s$^{-1}$ and $n_0=6.7\times 10^{18}$ m$^{-3}$. For BEC $B$, inter-atomic repulsion at the entrance to the etched rings is too strong to be overcome by the resonant injection mechanism described above because the mean-field energy at the entrance to, and inside, the rings far exceeds the energies of the lowest single-particle radial modes. Mean field repulsion therefore slows atoms that approach the ZP from the trailing edge of the BEC, making their incident velocity significantly less than $\overline{v}_x$. Consequently, the $f(\overline{v}_x)$ curve for BEC $B$ lies below those for both the single atom and BEC $A$ and reveals no resonances for the $\overline{v}_x$ values shown. As $\overline{v}_x$ increases, though, the incident kinetic energy begins to dominate the mean field energy, causing the $f(\overline{v}_x)$ curves for both BECs to approach those of a single atom.

We now consider how the peak density of the focused cloud, $n_f$, varies with $\overline{v}_x$ for quantum reflection of BEC $A$. In the inset to Fig. \ref{fig:focal}, the solid curve shows $n_f/n_0$ values determined from full numerical simulations of BEC $A$. The form of this curve can be understood by noting that $n_f/n_0$ is approximately proportional to $NR(\overline{v}_x)/n_0 l_x(l^{f}_{r})^2$, where $R(\overline{v}_x)$ is the fraction of atoms that quantum reflect from the ZP (dotted curve in Fig. \ref{fig:focal} inset). In Fig. \ref{fig:focal}, $n_f/n_0$ attains a peak value of $\sim 4$ when $\overline{v}_x =3$\:mm\:s$^{-1}$ $\thickapprox v_8$ \cite{footnote9}. For higher $\overline{v}_x$, $n_f/n_0$ decreases with increasing $\overline{v}_x$ because $R$ decreases rapidly, as expected from previous quantum reflection studies \cite{shimizu,pasquini,pasquini2,scott}. But as $\overline{v}_x$ decreases below $3$ mm s$^{-1}$, the atoms can no longer penetrate the narrow outer ZP rings, thus increasing $l^{f} _{r}$ and reducing $n_f/n_0$. Higher density focused clouds could be achieved either by fabricating fine (nm-scale) pillars within the raised ZP rings, to increase $R$ without affecting the diffraction process \cite{pasquini2}, or by using {\em{transmission}} ZPs, as we consider in the next section. 

\section{Transmission focusing and depletion of the BEC}
\label{sec:Transmission focusing and decoherence of the BEC}

\begin{figure}[t]
\includegraphics[width=1.0\columnwidth]{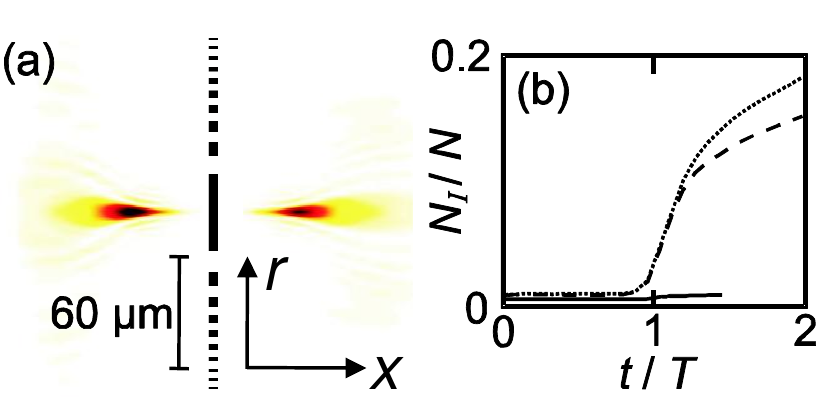} 
\caption{\label{fig:coher}(a) Atom density profile (dark high) showing transmission (right) and reflection (left) foci in the $x-r$ plane (axes inset) for BEC $A$ at $t$ = 104 ms. Black shapes: schematic cross-sections through solid regions of the ZP. Vertical bar indicates scale. (b) Fraction of incoherent atoms $N_I/N$ calculated versus $t$ for BEC $A$ (solid curve), BEC $B$ (dotted curve), and the reflected component \emph{only} of BEC $B$ (dashed curve). Calculations are for $\overline{v}_x=3$ mm\:s$^{-1}$.}
\end{figure}

In this section, we consider a transmission ZP, which has the same ring pattern as the etched plate considered in the previous section, but is only 130 nm-thick along the $x$ direction, as in the experiments of Ref. \cite{doak2}. Since the ring-shaped holes extend right through the plate, transmission ZPs are held together by a small number of radial struts, which do not affect the focusing process because their width is $\ll \lambda_{dB}$. In our calculations, we have no imaginary absorption potential in the gaps [white in Fig. \ref{fig:coher}(a)] so that all ($\sim N/2$) atoms entering the gaps emerge on the other side of the plate. Figure \ref{fig:coher}(a) shows the reflected (left) and transmitted (right) foci calculated for BEC $A$, which form at $t = 104$\:ms \cite{requals1}. The transmission focus contains $\sim N/2$ atoms, almost 50 times the number ($\sim R(\overline{v}_x)N/2$) in the reflected cloud in Fig. \ref{fig:basic}(d). Consequently, transmission ZP focusing can increase the density of the atom cloud passing through the plate by two orders of magnitude to $\sim 10^{20}$ m$^{-3}$. This compression increases the atom loss rate due to three-body scattering by six orders of magnitude \cite{pethick}. However, for BEC $A$, we estimate that the resulting fraction of atoms lost during the focusing process will be $< 0.1$. The radial width of the transmission focus in Fig. \ref{fig:coher}(a) is $\sim 5 \mu$m, suggesting that ZP focusing could assist the injection of BECs into the 10 $\mu$m-diameter hollow-core of a photonic crystal fibre \cite{christensen,hofferberth}, thereby increasing the fraction of atoms that can be transferred from a free-space trap into the fibre. 

Previous studies have shown that interactions between the incident and reflected components of a BEC that quantum reflects from a solid surface can partially decohere the atom cloud \cite{scottnoise}, particularly when its density is high. We have investigated whether the density increase produced by ZP focusing affects the coherence of BECs $A$ and $B$ by calculating the number of incoherent (i.e. non-condensed) atoms, $N_I$, as a function of time, $t$, using the truncated Wigner method described in Refs. \cite{norrieturb,decoherence}. For BEC $A$, the focusing process causes negligible incoherent scattering and depletion of the condensate, with the fraction of incoherent atoms, $N_I/N$ [solid curve in Fig. \ref{fig:coher}(b)], remaining below 0.01 throughout our simulation. By contrast, for BEC $B$ the incoherent fraction [dotted curve in Fig. \ref{fig:coher}(b)] rises sharply as the cloud strikes the ZP when $t=T=76$ ms. This is because BEC $B$ is $\sim 10$ times denser than BEC $A$, hence increasing the probability of incoherent scattering events \cite{scottnoise}. The rate of incoherent scattering, and consequent increase of $N_I/N$, is highest during the reflection process, i.e. when $1 \lesssim t/T \lesssim 1.2$. Thereafter, the rate decreases but remains finite due to inter-atomic scattering events that occur during the focusing process. In this regime, the contribution to $N_I/N$ made by atoms in the \emph{reflected} part of BEC $B$ only [dashed curve in Fig. \ref{fig:coher}(b)] deviates from the decoherent fraction of the \emph{whole} cloud [dotted curve in Fig. \ref{fig:coher}(b)]. However, since the deviation is very small, we conclude that quantum fluctuations deplete the reflected part of the condensate far more than the transmitted part. Physically, this is because collisions between the incident and reflected matter waves give rise to incoherent scattering processes that do not affect the transmitted wave \cite{requals1}. Since transmission focusing does not significantly deplete the condensate, it may provide a useful tool for manipulating BECs, for example to transfer them into microtraps or hollow core optical fibres or to co-focus coherent light and matter waves to ensure strong interaction between them \cite{hofferberth}.   

As an alternative to using the magnetic trap displacement technique described in Section \ref{sec:Quantum reflection from the zone plate} to direct the BEC towards the ZP \cite{pasquini,pasquini2} a moving optical lattice, formed by two counter-propagating laser beams with slight relative frequency detuning, would reduce the velocity spread of the incident atoms and the associated chromatic aberration. This technique might also allow ZP focusing of 2D atom clouds \cite{ChipOptLatt}, which can be trapped within each lattice minimum and passed sequentially through the ZP. This would combine high flux, sufficiently low densities to reduce potentially harmful inter-atomic interaction effects, and good focusing properties due to the small transverse velocity spread given by the (single particle) ground state momentum distribution within the individual wells.

\section{Zone-plate lithography of two-dimensional electron gases}
\label{sec:Zone-plate lithography of two-dimensional electron gases}
\subsection{Effect of adsorbed atoms on the electron gas}
We now explore the possibility of using ZPs controllably to deposit alkali atoms onto a semiconductor surface oriented parallel to the plate and in its focal plane, so enabling matter-wave lithography of quantum electronic components such as quantum wires and dots within a 2DEG.

When alkali atoms are deposited on materials with a higher electronegativity, they polarize by the partial transfer of their valence electron to the surface \cite{polarize,polarize2}. Stronger polarization is expected for heavier, less electronegative, alkali atoms. For example, polarization of a single Rb atom on a Si surface creates an electric dipole of magnitude $\sim 10^{-29}$ Cm pointing away from the surface \cite{polarize}. In this section, we consider Rb atoms because they are highly polarizable and hence have low electronegativity \cite{atomtype}. The interaction of alkali atoms with GaAs surfaces has also been extensively studied and continues to attract considerable interest, partly because it provides a way to lower the work function of GaAs, which is important for technological applications in Schottky barriers \cite{GaAs1,GaAs2,GaAs3,GaAs4,GaAs5,GaAs6,GaAs7,footCs}. Polarization of the adsorbed atoms creates an electric field, which, as we now explain, can strongly affect the density and electrical resistance of a two-dimensional electron gas (2DEG) just below the surface.

In a 2DEG, electrons from remote ionized donors form a sheet of negative mobile charge, $\sim 15$ nm thick, parallel to the surface plane \cite{2DEG}. Typically, the 2DEG is located a few 10s of nm below the surface, although it can be on the surface itself \cite{2DEGonsurface}. A voltage, applied to Ohmic contacts, creates an electric field along the 2DEG, thus driving current. Since the electrons are spatially separated from the parent ionized donors, their mobility is usually very high, particularly at low temperatures. Consequently, 2DEGs are used extensively in condensed matter research and also have applications in high-frequency electronics: mobile telephones, for example. A 2DEG can be located within $\sim 50$ nm of the semiconductor surface \cite{qwire,2DEGonsurface}, which is close enough for the potential energy due to repulsion betwen electrons in the 2DEG and the dipoles created by the adsorbed alkali atoms to be 10s of meV. This is sufficient locally to deplete the 2DEG, whose Fermi energy is typically $\sim 10$ meV, so producing a large measurable increase in the 2DEG's resistance.

\begin{figure}
\includegraphics[width=1.0\columnwidth]{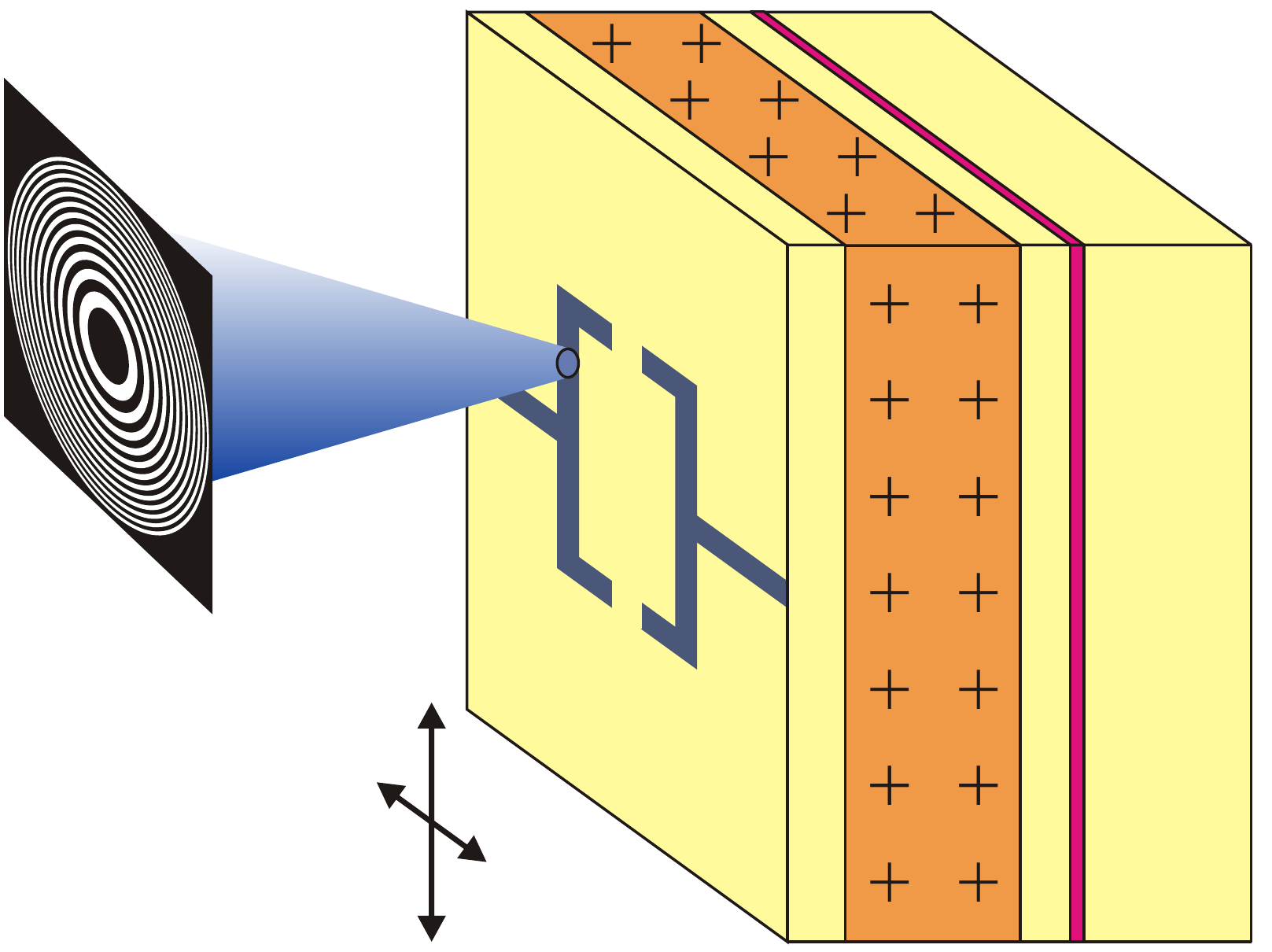} 
\caption{\label{fig:ZPschema}Schematic diagram of the heterostructure showing the GaAs layers (light yellow), (AlGa)As layer (dark yellow), the two $\delta$-doping layers (+) and the 2DEG (red). Shaded blue cone represents the focusing and surface deposition of $^{87}$Rb atoms after they have passed through the ZP (black and white structure on the left-hand side of the figure). Scanning the heterostructure in the directions arrowed, parallel to a single ZP or a ZP arrray \cite{MITarray,MITarray2}, deposits the atoms in surface patterns with arbitrary shapes. Polarization of atoms within the surface pattern imprints a similarly-shaped depletion region in the 2DEG. For illustration, here the atoms are deposited within the blue regions on the surface to create a square quantum dot in the 2DEG, like that studied in \cite{marlow}. Electrons enter and leave the dot via quantum point contact openings created by the gaps in the upper and lower edges of the blue square. The two arms emerging from the left- and right-hand edges of the blue square form depletion barrriers in the 2DEG, which prevent electrons flowing around the outside of the dot \cite{marlow}.}
\end{figure}

To illustrate this, we have investigated the effect of $^{87}$Rb atoms deposited, by ZP focusing of a BEC, onto the surface of a GaAs/(AlGa)As heterostructure containing a 2DEG at a distance $l=$ 42 nm below the surface \cite{atomtype,2DEGonsurface}. We expect similar results for 2DEGs in Si-based devices. Figure \ref{fig:ZPschema} shows a schematic diagram of the heterostructure and ZP-lithography process. The 2DEG is formed by two $\delta$-doping layers, located 22 and 32 nm below the surface and of density $1.3 \times 10^{16}\text{m}^{-2}$ and $10^{16}\text{m}^{-2}$ respectively, similar to values used in recent experiments \cite{taylor,ourphysicaE}. For this sample, self-consistent solution of Poisson's equation perpendicular to the surface shows that the Fermi energy of the 2DEG is $E_F=10^3 \pi \hbar^2 n_{2DEG}/e m^*$ = 2.9 meV, where $n_{2DEG}= 8 \times 10^{14} \text{m}^{-2}$ is the sheet electron density, $e$ is the magnitude of the electronic charge, and the electron effective mass, $m^*$, is 0.067 times the free electron mass \cite{2DEG,ourphysicaE}. 

To deposit the atoms and measure their effect on the 2DEG, the BEC and heterostructure must be held in an ultra-high vacuum system with a background pressure in the low $10^{-11}\text{mbar}$ range, as in the surface-physics experiments of Refs. \cite {GaAs1,GaAs2,GaAs3,GaAs6}, for example. Since precisely the same vacuum condition is also used in standard BEC experiments, the environments required to produce and controllably deposit the ultracold atoms are compatible. We consider atoms deposited onto the Ga-rich surface of GaAs (001)-(4 $\times$ 2)/$c$(8 $\times$ 2), which produces strong ionic bonding of the alkali-atom adsorbates due to partial electron transfer from the atom to the substrate \cite{GaAs1,GaAs4,GaAs7}. This Ga-rich surface reconstruction can be prepared by encapsulating the heterostructure, after molecular beam epitaxial growth, with an As overlayer and then removing this overlayer within the vacuum system by annealing the heterostructure at temperatures up to 850 K \cite {GaAs1,GaAs3}. To heat the heterostructure to such high temperatures, and then cool it to 85 K or below so that the adsorbed atoms do not diffuse \cite{temp_diffusion1,GaAs2,GaAs3}, requires a sample holder like that described in Refs. \cite{GaAs3,GaAs8}, which allows the temperature to be stabilized at any value between 85 K and 850 K. 

Analysis of the BEC's focusing dynamics, obtained by solving Eq. (\ref{eq:gpe}) numerically, shows that, to good approximation, the density profile of atoms deposited on the surface is of the form $n_{surf}(r)=n_{P} \exp(-r^2/ \lambda^2)$, where the peak density $n_P=N_{atom}/(2 \pi \lambda^2)$ increases with the total number of atoms deposited, $N_{atom}$. We consider $n_{P} \geqslant 10^{17} \text{m}^{-2}$ (the corresponding atom numbers are specified below), so that the mean inter-atomic spacing ($\lesssim 3$ nm) is much less than the distance (42 nm) between the surface and the 2DEG. This ensures that the atoms can be modelled by the continuous distribution $n_{surf}(r)$, when calculating their effect on the 2DEG. Since Ga has a similar electronegativity to that of Si, we take the electric dipole moment of each adsorbed atom to equal that measured previously ($10^{-29}$ Cm) for $^{87}$Rb on Si \cite{polarize}, which is comparable to values calculated using density-functional-theory for alkali atoms on GaAs \cite{GaAs4,GaAs7}. We note, however, that our results do not depend critically on the precise value of this parameter. To determine how the atoms influence the 2DEG, we calculated the electron potential energy within and above the heterostructure by solving Poisson's equation in cylindrical co-ordinates. Our calculations used a relaxation method with a variable mesh to capture the vastly different length scales that characterize the potential variation near and away from the surface dipoles. We included the effects of the adsorbed surface atoms, electron surface states, shallow donor states, and linear (Thomas-Fermi) screening by the 2DEG \cite{ourphysicaE}. Our calculations use a ``frozen charge'' model \cite{davies}, in which the adsorbed atoms change the potential within the heterostructure and the electron density profile in the 2DEG, but do not alter the distribution of charge within the mid-gap surface states or donor layers.

\begin{figure}[t]
\includegraphics[width=0.72\columnwidth]{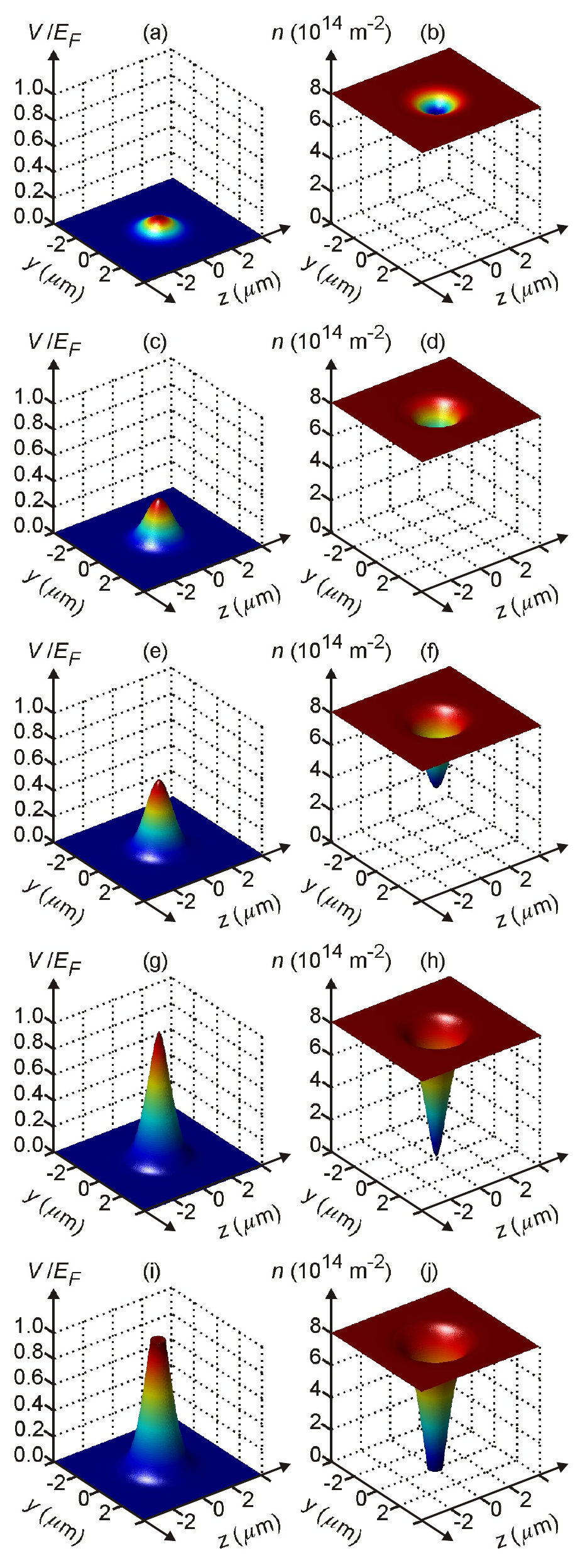} 
\caption{\label{fig:2DEG}Electron potential energy, $V(r=\sqrt{y^2+z^2})$, (left-hand column) and density, $n(r)$, (right-hand column) in the plane of the 2DEG directly below $10^{6}$ [(a) and (b)], $3 \times 10^{6}$ [(c) and (d)], $5 \times 10^{6}$ [(e) and (f)], $8 \times 10^{6}$ [(g) and (h)], $10^{7}$ [(i) and (j)] $^{87}$Rb atoms deposited, with a Gaussian profile of width $\lambda = 1$ $\mu$m, on the surface of the heterostructure, which lies in the focal plane of a transmission ZP. In (j), the 2DEG is fully depleted by its repulsive interaction with the polarized surface atoms.}
\end{figure}
 
We first consider atoms deposited with a radial spread $\lambda = 1$ $\mu$m consistent with the focal width shown in Fig. \ref{fig:coher}(a). Figure \ref{fig:2DEG}(a) shows the electron potential energy variation, $V(r)$, in the plane of the 2DEG directly below $N_{atom}= 10^6$ adsorbed atoms. Repulsive interaction with the polarized atoms increases the electron potential energy by $\sim E_F/10$ when $r=0$. Figure \ref{fig:2DEG}(b) shows the corresponding electron density, $n(r)=[E_F-V(r)]e m^* /10^3 \pi \hbar^2$, which decreases to approximately 90\% of its bulk value as $r \rightarrow 0$. Figure \ref{fig:2DEG} reveals that as $N_{atom}$ increases to $3 \times 10^{6}$ [(c) and (d)], $5 \times 10^{6}$ [(e) and (f)], $8 \times 10^{6}$ [(g) and (h)], the peak value of $V(r)$ gradually increases to $E_F$ and, consequently, $n(r)$ falls to zero below the centre of the surface atoms. When $N_{atom} = 10^{7}$ [Fig. \ref{fig:2DEG}(i) and (j)], which requires several different BECs to be deposited, sequentially, on the surface, the 2DEG is fully depleted when $r \lesssim 0.5$ $\mu$m and $V(r) > E_F$. For a range of $\lambda$ values, we find that total depletion needs only a low density of adsorbed atoms, $n_P \approx 10^{18}\text{m}^{-2}$, corresponding to approximately 0.1 monolayers, which ensures that these atoms interact far more strongly with the surface than with one another \cite{GaAs3,GaAs4}.

Experimental confirmation of the local electron depletion could be obtained by depositing the atoms immediately above a quantum wire, $\sim$ 1 $\mu$m across and 5 $\mu$m long, microfabricated in the 2DEG \cite{sachradja,crook}. As the atoms are deposited, the resistance of the quantum wire would rapidly increase, as observed previously when circular antidots (depletion regions) are introduced in a narrow conducting channel \cite{sachradja,crook,lockin}. Alternatively, to determine the profile of the adsorbed atoms spatially and as a function of time, quantum wires \cite{qwire}, each comprising a quasi one-dimensional (1D) conduction channel, could be fabricated within the 2DEG by implanting ions into the semiconductor material, for example Ga ions in GaAs, thus locally disrupting the 2DEG and transforming it from a conductor to an insulator \cite{ion_beam1,ion_beam2,ion_beam3,ion_beam4,ion_beam5,ion_beam6}. Ion beam implantation can define a conduction network comprising two arrays of quantum wires, each containing narrow ($\sim 0.1-1$ $\mu$m) parallel conduction channels, which intersect at right angles. Monitoring the resistance of each quantum wire, would enable the deposition of alkali atoms on the surface of the device, or even held above it \cite{lockin}, to be mapped as a function of space and time, with sub-micron spatial resolution determined by the width of the wire. Overcoming the $\sim \mu$m resolution limit of optical imaging is important for a wide range of ultracold-atom experiments including studies of tailored interacting many-body systems \cite{manybodyrev} where correlation functions \cite{HBTAspect} could be measured with unprecedented spatial resolution, \emph{in situ} observation of soliton and vortex creation and dynamics \cite{vortexdyn}, and the study of atom-surface interactions \cite{KrugerPRA}.

The behaviour of adsorbed atoms depends on the temperature of the semiconductor surface \cite{temp_diffusion1,GaAs2,GaAs3}. At room temperature, the atoms will diffuse across the surface at a rate that could be determined by measuring the time evolution of the quantum wires' resistance. By contrast, below 85 K, the atoms will stick where they are deposited by the ZP \cite{GaAs2,GaAs3}, thus producing a well-defined surface polarization pattern and electric field profile. The spatial resolution of such patterns is limited by the radial width, $l^{f}_{r}$, of the focused cloud, which, as discussed in Section \ref{sec:Quantum reflection from the zone plate}, approximately equals the width of the narrowest ZP ring that the matter wave passes through \cite{meyer,ZP_width_theory,focusing,effective}. A BEC with strong repulsive interactions is unable to penetrate ZP rings narrower than the healing length (typically a few hundred nm in a trapped BEC), which therefore limits $l^{f}_{r}$. To investigate whether this limitation can be overcome by suppressing inter-atomic interactions, we have studied the focusing and deposition of an atom cloud that is trapped optically and subject to a small uniform magnetic field tuned to a Feshbach resonance so that $a \approx 0$ in Eq. (\ref{eq:gpe}), as can be achieved for a range of alkali atoms \cite{optical_move,width,attractive}.

\begin{figure}[t]
\includegraphics[width=1.\columnwidth]{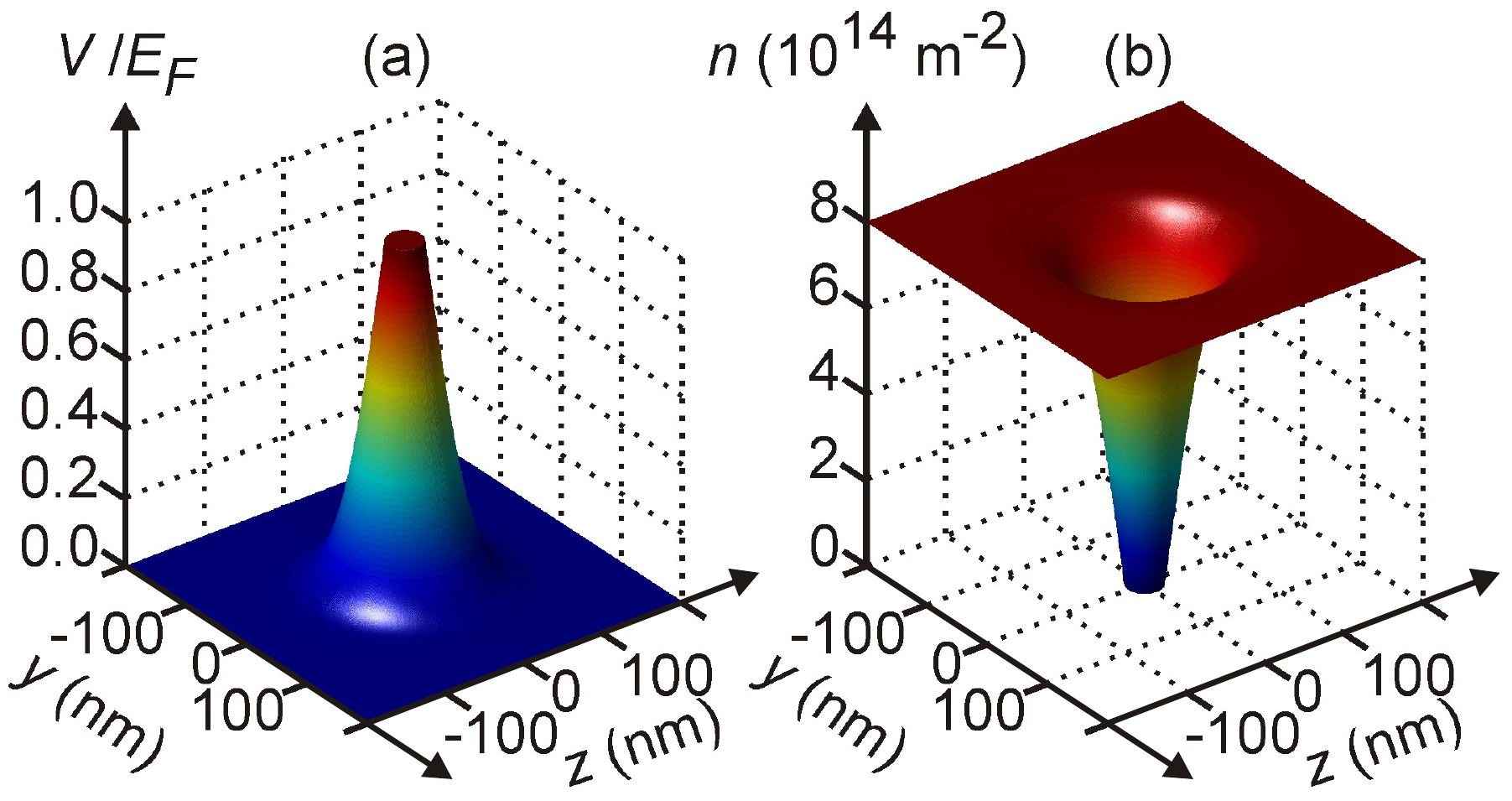} 
\caption{\label{fig:B}(a) Electron potential energy, $V(r=\sqrt{y^2+z^2})$, and (b) electron density, $n(r)$, calculated in the plane of the 2DEG directly below 1800 $^{87}$Rb atoms focused, with a Gaussian profile of width $\lambda = 25$ nm, onto the surface of the heterostructure. The 2DEG is fully depleted by its repulsive interaction with the polarized surface atoms.}
\end{figure}

When there are no inter-atomic interactions, $l^{f}_{r}$ is limited only by the narrowest ZP ring that the atom cloud spans \cite{ZP_width_theory,ourcalcs,focusing}, which can be etched as small as 12 nm \cite{ZP2009}, with expectations that ring widths $<$ 10 nm can be achieved \cite{ZP_width_theory,ZP10nmA,ZP10nmB,ZP10nmC,ZP2009,attractive}. Consequently, the width of the electron depletion region produced by the adsorbed atoms is limited either by $l^{f}_{r}$ or by the distance from the 2DEG to the surface, whichever is the larger. In both semiconductor heterostructures and graphene, 2DEGs can form on the surface itself \cite{2DEG,geim}, suggesting that depeletion regions a few 10s of nm across are attainable using existing ZPs \cite{ZP2009}. For the heterostructure considered here, though, adsorbed alkali atoms will not produce depletion regions smaller than the 2DEG-surface separation $l=$ 42 nm. We therefore study atoms focused by the ZP reported in Ref. \cite{doak2}, which has $R_1$ = 10.4 $\mu$m and an outer ring width of 50 nm, small enough to enable alkali-atom lithography of quantum components in 2DEGs. To demonstrate this, we calculated $V(r)$ in the plane of the 2DEG, taking $\lambda$ = 25 nm, corresponding to a full focal width of $\sim$ 50 nm. After the deposition of only 1800 atoms, when $r \lesssim$ 50 nm, $V(r) > E_F$ [Fig. \ref{fig:B}(a)] and the electron density is zero [Fig. \ref{fig:B}(b)]. Numerical solutions of Eq. (\ref{eq:gpe}) reveal that this sharp focus can be obtained for a range of approach speeds, for example $\overline{v}_x=2.4$ mm\:s$^{-1}$, for which $f = 57 \mu$m. Due to this long focal length, the ZP is positioned far from the surface, thus overcoming a major problem of mask-based lithography, namely that to produce features on a 10 nm scale the mask must be held within $\sim 1$ $\mu$m of the surface \cite{MITarray2}, which is extremely challenging due to the strong Casimir attraction at such small separations.

\subsection{Prospects for ZP-based matter-wave lithography}

We now consider routes to exploiting ZP-focusing of matter waves for flexible, high-speed, erasable lithography of quantum electronic components.

It is not necessary for ZPs to be circular in order to focus incident waves. For example, 1D ZPs comprising diffraction slits with edges at positions $R_n=R_1 n^{1/2} (n=1,2,3,...)$ can focus waves into a narrow line \cite{alloschery}. Consequently, focusing matter waves by 1D and/or circular ZPs can create a wide range of surface patterns, thus complementing existing matter-wave lithography techniques, such as the use of optical standing waves to focus and deposit atoms in parallel lines \cite{McClelland,Timp1,Timp2,ODwyer,pfau}. Alkali-atom lithography using ZPs offers comparable resolution (a few 10s of nm) to that demonstrated with optical lattices \cite{pfau}, plus the flexibility to tailor the distribution of adsorbed atoms using robust and well developed ZP systems. Moreover, in contrast to ion implantation \cite{ion_beam1,ion_beam2,ion_beam3,ion_beam4,ion_beam5,ion_beam6}, which damages the heterostructure and degrades the electron mobility, alkali-atom lithography of 2DEGs is a gentle non-invasive process that requires no dedicated mask to produce each desired surface pattern. 

By mounting the heterostructure on a scanning stage and moving it under the ZP with nm-precision (see Fig. \ref{fig:ZPschema}), arbitrary patterns of adsorbed atoms (blue areas on surface in Fig. \ref{fig:ZPschema}) and resulting electron depletion could be written in dot-matrix fashion \cite{MITarray,MITarray2}. Planar \emph{arrays} of transmission ZPs developed at MIT for maskless X-ray lithography \cite{MITarray,MITarray2} are also suitable for depositing many alkali atom spots \emph{in parallel} on a surface. Such ZP-array lithography has major advantages over other techniques: fast write speed, scalability, and the flexibility to produce a wide range of surface patterns including arrays of identical components, lines, gratings, and interconnects. In the case of matter waves, cold atoms would be supplied to each ZP either by outcoupling them from a single BEC, as in an atom laser \cite{atomlaser1,atomlaser2,atomlaser3,atomlaser4,atomnumber}, or by using an array of microtraps like that created recently by using permanent magnets to confine the atoms \cite{hall,fernholz1,fernholz2,otherways}, with each trap supplying atoms to a particular ZP.  

The polarization of the adsorbed atoms can be controlled \emph{in situ} by applying an electric field perpendicular to the surface \cite{polarize,kruger_efield}. This suggests a way to alter the quantum components, or temporarily erase them, by using an electric field directed towards the surface to depolarize the adsorbed atoms. Surface patterns, or even individual components, could be permanently erased by using optical or UV radiation to either remove the atoms from the surface or make them diffuse away due to local heating \cite{polarize,polarize2,desorb1,desorb2,desorb3,desorb4,desorb5,desorb6,desorb7,desorb8,desorb9,desorb10}, and then re-written on the same wafer. Erasable lithography of quantum components in a 2DEG is of considerable interest because it allows individual components to be modified or repaired \cite{crook}. Moreover, the capacity to fabricate different devices sequentially, on the same surface, is crucial for distinguishing the effects of device \emph{geometry} on quantum transport from those originating from \emph{material} impurities and imperfections \cite{taylor,micolich1,micolich2,crook,crookref12}. So far, though, erasable lithography has only been achieved for single quantum electronic components by using scanning probe techniques \cite{crook,crookref12}, which produce charge patterns on the surface. Compared with such techniques, ZP-based alkali-atom lithography offers high write speed and scalability. It may also be surprisingly cost-effective because BECs can now be created using off-the-shelf kits costing less than \$50K \cite{cheapaschips}.

Polarization of dense patches of adsorbed alkali atoms creates a strong inhomogeneous electric field \emph{above} the surface of the heterostructure as well as below it \cite{polarize,polarize2,kruger_efield}. Consequently, ultracold atoms above the surface and electrons within the 2DEG would move in similarly-shaped potentials. Their motion may therefore correlate, or even couple, suggesting a route to developing hybrid electronic/atomic microchip structures made by ZP lithography. Since such structures are potentially re-writable, they could be fabricated and studied \emph{in situ} without needing to break the vacuum between experiments on different chip geometries -- thus greatly reducing the time between device fabrication and measurement.

\section{Summary}
\label{sec:Summary}
In summary, BECs can be sharply focused by quantum reflection from, or transmission through, Fresnel ZPs. Optimal focusing, achieved for flat dilute BECs with $l_x/l_r \ll 1$ and $n_0<10^{18}$ m$^{-3}$, can increase the peak atom density by up to two orders of magnitude. Despite the increased atom-atom scattering rates that accompany this compression, the focusing process does not significantly reduce the condensed fraction of the atom cloud. Focal lengths obtained from numerical simulations of the Gross-Pitaevskii equation are similar to those expected from a single-particle ray analysis, but exhibit additional resonances originating from inter-atomic interactions. Transmission ZP focusing of matter waves provides a powerful lithographic tool for fabricating quantum electronic components by depositing well-defined, potentially re-writable, patterns of atoms on the surface of a semiconductor heterojunction containing a 2DEG \cite{footnote2DEG}. This new type of lithography offers state-of-the-art resolution, scalability by using ZP arrays, the ability to re-write all or selected components, and a possible route to creating hybrid electron/atom chips that are fabricated and studied \emph{in situ}. Since all of the individual components required to realize ZP-based alkali-atom lithography exist, we hope that our results will stimulate the experimental work required to unite these components in a practical demonstration of the technique. 

\section*{Acknowledgements}
This work is funded by EPSRC UK and the ARC. We thank Lucia Hackerm{\"u}ller and Philip Moriarty for helpful discussions.

\bibliography{zpnewbib}

\begin{thebibliography}{150}
\bibitem[{lig()}]{lightfocus}
\bibinfo{note}{The light and matter waves have a common focal point whenever
  their wavelengths are equal.}
\bibitem[{foo({\natexlab{a}})}]{footnote8}
\bibinfo{note}{Since all atoms entering the ring are adsorbed by the imaginary
  potential, our results are insensitive to the details of the real part of the
  potential within the ring.}
\bibitem[{foo({\natexlab{b}})}]{footnote7}
\bibinfo{note}{There is only one focal point, where the density of the
  reflected cloud is maximal, because the wave amplitude varies almost
  sinusoidally across the etched rings \cite{meyer}.}
\bibitem[{opt({\natexlab{a}})}]{opticalZP}
\bibinfo{note}{Scanning laser beams have been used previously to define optical
  billiards for cold atoms \cite{optical_billiard1,optical_billiard2}. Our
  calculations predict that using this technique to phase imprint the ZP
  pattern causes the BEC to collapse rapidly into a focused spatial region
  whose size is comparable to the narrowest ZP ring written by the beam -- in
  practice limited by the width of the beam itself.}
\bibitem[{foo({\natexlab{c}})}]{footnote10}
\bibinfo{note}{Since the energies of the quantized modes in the rings will be
  increased by inter-atomic repulsion, the threshold velocity required for
  resonant injection into each ring will be slightly higher than expected from
  this simple single-particle analysis.}
\bibitem[{foo({\natexlab{d}})}]{footnote2}
\bibinfo{note}{We only consider $\Delta x > l_x/2$ ($\overline{v}_x > 2$ mm
  s$^{-1}$) so that the initial wavepackets do not touch the ZP surface.}
\bibitem[{req()}]{requals1}
\bibinfo{note}{To obtain an upper bound on the fraction of thermal atoms
  created during the focusing process, in this section we take $R=1$ so that
  there is maximal interaction between the incident and reflected parts of the
  matter wave. When $R<1$, the fraction of thermal atoms is lower in the
  reflected cloud but unchanged in the transmitted cloud.}
\bibitem[{dec()}]{decoherence}
\bibinfo{note}{We do not consider decoherence due to interaction between the
  atoms and the surface, which does not significantly degrade diffraction
  patterns observed for individual atoms or molecules passing through gratings
  with slits that are only a few 10s of nm wide
  \cite{carnal,doak,doak2,zeilinger,hack1,hack2,hack3,hack4,hack5}. There have
  been few studies of the dynamics of BECs passing through, or trapped within,
  solids. However, for a BEC trapped inside a hollow-core photonic crystal
  fibre \cite{christensen}, interactions with the surface produce little
  heating of the trapped atom cloud, which can have a lifetime in excess of 50
  ms -- far longer than the time (10s of $\mu$s) taken for an atom to pass
  through the transmission ZP considered here.}
\bibitem[{ato({\natexlab{a}})}]{atomtype}
\bibinfo{note}{We consider $^{87}$Rb atoms because their polarizability is
  higher than $^{23}$Na. However, $^{23}$Na atoms will have a similar effect on
  the 2DEG if an applied electric field is used to enhance their electric
  dipole moment \cite{polarize,kruger_efield}. Maximal 2DEG depletion would be
  obtained by using such a field in combination with a heavy alkali atom, like
  $^{133}$Cs \cite{footCs}, with an intrinsically high polarizability.
  $^{133}$Cs also offers other advantages due to its stability on GaAs
  \cite{footCs}, but is less frequently used to create and study BEC than
  $^{23}$Na or $^{87}$Rb.}
\bibitem[{foo({\natexlab{f}})}]{footCs}
\bibinfo{note}{Most studies of alkali metal adsorption on GaAs have focused on
  $^{133}$Cs, usually deposited on the Ga-rich surface \cite{GaAs5}, because it
  does not form chemical compounds with the semiconductor material or diffuse
  into the bulk and is less prone to clustering than $^{23}$Na, $^{39}$K, or
  $^{87}$Rb \cite{GaAs5,GaAs6}.}
\bibitem[{2DE()}]{2DEGonsurface}
\bibinfo{note}{In graphene \cite{geim} and inversion layers on InAs-based
  heterostructures \cite{2DEG}, the 2DEG is on the surface of the sample and
  therefore particularly sensitive to adsorbed atoms and molecules
  \cite{gasongraphene}. There is considerable interest in K adsorption on
  graphene, which, as for alkali atoms on Si or GaAs, also produces a high
  surface electric dipole \cite{Kongraphene}.}
\bibitem[{foc()}]{focusing}
\bibinfo{note}{As atoms are adsorbed, they will polarize and hence create an
  electric field above the surface. This field will attract those atoms that
  are still approaching the surface \cite{polarize,kruger_efield}, and
  therefore tend to reduce the focal width. Consequently, the distribution of
  atoms deposited on the surface may be narrower than the width of the
  outermost ZP ring.}
\bibitem[{eff()}]{effective}
\bibinfo{note}{Due to atom-surface van der Waals interactions, the effective
  width of the slit is reduced with respect to its physical size
  \cite{grisenti,zhou}.}
\bibitem[{opt({\natexlab{b}})}]{optical_move}
\bibinfo{note}{The optical trap can be moved towards the surface at speeds of a
  few $\text{mm s}^{-1}$ by displacing the laser beams that produce confinement
  along the $x-$direction \cite{cornish}.}
\bibitem[{wid()}]{width}
\bibinfo{note}{Alternatively, atoms could be deposited in fine patterns on the
  surface by holding them in a tightly-confined one-dimensional Bose gas on an
  atom chip \cite{schmiedmayer1D} or in an optical lattice \cite{moritz}, and
  then displacing the trap centre so that it lies within the surface.}
\bibitem[{att()}]{attractive}
\bibinfo{note}{By exploiting the Feshbach resonance, it may be possible to make
  $l^{f}_{r}$ significantly smaller than the narrowest ring width by changing
  the magnetic bias field when the BEC focuses so that $a<0$ and the cloud
  shrinks due to the attractive inter-atomic interactions.}
\bibitem[{our()}]{ourcalcs}
\bibinfo{note}{Our numerical solutions of Eq. (\ref{eq:gpe}) with $a=0$,
  confirm that, in the absence of inter-atomic interactions, the radial width
  of the focused atom cloud is very similar to the width of the outermost ZP
  ring.}
\bibitem[{foo({\natexlab{g}})}]{footnote2DEG}
\bibinfo{note}{2DEGs, cooled to liquid helium temperatures, could offer several
  advantages over metal surface wires for creating the magnetic trapping
  fields. Firstly, the higher electrical resistance of the 2DEG would reduce
  Johnson noise \cite{lin, henkelnoisy,harber,jonessurf,zhang}, whilst its high
  mobility would suppress spatial fluctuations in the current flow and magnetic
  field profile. Secondly, the extremely low thickness of the 2DEG sheet ($\sim
  15$ nm), combined with high-precision semiconductor fabrication techniques,
  would reduce both edge fluctuations \cite{esteve} and spin flip rates
  \cite{lin, henkelnoisy,harber,jonessurf,zhang}, thus potentially allowing
  atoms to be trapped closer to the surface. Finally, using ion implantation to
  define the quantum wires would leave the chip surface flat and intact, thus
  allowing additional components such as voltage gates or wires to be
  fabricated on the surface \cite{kruger_efield}.}
\end{thebibliography}

\end{document}